# Graphene-based quantum capacitance wireless vapor sensors

David A. Deen, *Member, IEEE*, Eric J. Olson, Mona A. Ebrish, *Student Member, IEEE,* and Steven J. Koester, *Senior Member, IEEE*

*Abstract*—A wireless vapor sensor based upon the quantum capacitance effect in graphene is demonstrated. The sensor consists of a metal-oxide-graphene variable capacitor (varactor) coupled to an inductor, creating a resonant oscillator circuit. The resonant frequency is found to shift in proportion to water vapor concentration for relative humidity (RH) values ranging from 1% to 97% with a linear frequency shift of 5.7 $\pm$ 0.3 kHz / RH%. The capacitance values extracted from the wireless measurements agree with those determined from capacitance-voltage measurements, providing strong evidence that the sensing arises from the variable quantum capacitance in graphene. These results represent a new sensor transduction mechanism and pave the way for graphene quantum capacitance sensors to be studied for a wide range of chemical and biological sensing applications.

## I. INTRODUCTION

THE quantum capacitance effect is a direct, observable manifestation of the Pauli exclusion principle. While this effect is particularly prominent in the two-dimensional material graphene [1-13] due to its low density of states, few if any practical uses for this effect have been demonstrated to date. It has recently been proposed that the quantum capacitance effect could be utilized to realize wireless sensors due to graphene's energy-dependent density of states and excellent surface sensitivity [14]. Such a device could have significant advantages over alternative techniques, such as resistance-based sensing [15-21] and wireless sensing based upon microelectromechanical systems [22,23]. Here we demonstrate graphene-based wireless vapor sensors that utilize the variable capacitance that arises due to the energy-dependent density of states as the sensor transduction mechanism. Graphene variable capacitors (varactors) are

Manuscript received ________, 2013. This work was supported by the Minnesota Partnership for Biotechnology and Medical Genomics Decade of Discovery Initiative. This work also utilized the University of Minnesota Nanofabrication and Characterization Facilities, which receive partial support from the National Science Foundation.

D. A. Deen, E. J. Olson, M. A. Ebrish, and S. J. Koester are with the Department of Electrical and Computer Engineering, University of Minnesota, 200 Union St. SE, Minneapolis, MN 55455 (e-mail: dadeen@umn.edu; olso4499@umn.edu; ebris001@umn.edu; skoester@umn.edu). D. A. Deen is now with Seagate in Bloomington, MN (email: david.deen@alumni.nd.edu)

coupled to an inductor coil whereby the resonant frequency of the resulting LRC circuit shifts in response to the $H_2O$ vapor concentration, as determined using a secondary readout inductor [24]. We show strong evidence that the frequency shift arises from changes in the quantum capacitance in graphene, and that the resonant frequency shift shows a monotonic dependence on vapor concentration over a wide relative humidity range of 1% to 97%. Moreover, the response is shown to be reversible and stable upon repeated concentration cycling. The response time of the sensors was characterized and found to be comparable to the temporal resolution of the measurement setup. The advantages of graphene quantum capacitance wireless sensors compared to alternative passive sensing approaches include excellent noise immunity, greatly improved size scalability, fast response and potential for sensing a wide range of species depending upon the surface functionalization utilized. Our results suggest that graphene quantum capacitance wireless sensors can enable a powerful platform for detection of a wide range of chemical and biological targets [21, 25-29].

The basic transduction mechanism for the sensors utilized in this work is shown conceptually in Fig. 1. A change in the concentration, $\Delta M$, of adsorbed molecules on the graphene surface can change the carrier concentration in the graphene, $\Delta n$. Due to the low density of states in graphene, this leads to a measureable shift in the Fermi energy, $\Delta E_F$, as well as the quantum capacitance, $\Delta C_Q$. If the graphene is used as the electrode in a metal-graphene-oxide capacitor and this capacitor is integrated with an inductor, changes in the quantum capacitance lead to a resonant frequency shift, $\Delta f$, of the resulting LRC resonator circuit.

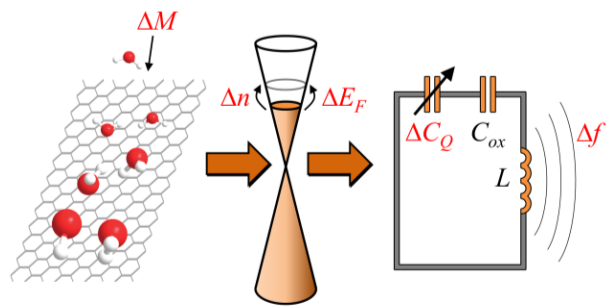

FIG. 1. Conceptual illustration of the graphene quantum capacitance vapor sensing transduction mechanism.

In order for this transduction mechanism to be utilized for gas sensing, the graphene must be exposed to the external environment, suggesting an inverted capacitor geometry with the graphene on top of the metal gate electrode. In addition, the capacitor dielectric must be sufficiently thin so that the quantum capacitance can significantly affect the overall capacitance of the system. Finally, the resonator must have high quality factor, $Q$, suggesting a multi-finger geometry in order to reduce the series resistance. We note that the transduction mechanism illustrated in Fig. 1 is fundamentally different than the graphene-based wireless sensor demonstrated in reference [21] where the resistance change of the graphene functionalized to be sensitive to bacteria was used to change the $Q$ of an LRC circuit rather than the resonant frequency.

## II. METHODS

### A. Device Fabrication

The graphene varactors were fabricated by first preparing a substrate by depositing $Si_3N_4$ followed by $SiO_2$ by plasma-enhanced CVD on a quartz substrate. The insulating quartz substrate minimizes parasitic capacitances associated with contact pads during high frequency measurements. Device processing relied upon conventional photolithography techniques and was initiated by a reactive ion recess etch of the $SiO_2$ layer and subsequent electron-beam deposition of the local back-gate metal (Ti/Pd). An 8-nm-thick $HfO_2$ layer was deposited by ALD for gate insulation and vias were patterned and dry etched through the $HfO_2$ layer to allow access to the gate pad. CVD-grown graphene was then transferred onto the patterned wafer. The single-layer graphene was grown on a copper foil, and spin-coated with PMMA. After baking, the graphene on the uncoated side of the foil was removed using an $O_2$ plasma etch. Next, the Cu was removed by using a $FeCl_3$-based etch and rinsed multiple times in HCl and deionized water. Finally the graphene layer attached to the PMMA, was transferred onto a substrate using an aqueous transfer process and the PMMA removed using a solvent etch. The graphene was then patterned using an $O_2$ plasma to define the desired active device geometries. Ohmic contacts were formed by electron-beam evaporation of a Ti/Pd/Au (1 nm / 25 nm / 35 nm) metal stack. Finally, thick Ti/Al (10 nm / 300 nm) pad metallization was deposited to allow bond wires to be attached to the devices. Following device fabrication, the presence of single-layer graphene was verified using Raman spectroscopy.

The final chip had numerous devices. All varactors had gate length of 5 μm and were arranged in multi-finger geometries, with finger length of either 40 μm or 100 μm. The multi-finger design allows large capacitances to be obtained while maintaining low series resistance. This graphene-on-top geometry has the additional advantage that it allows the dielectric to be made extremely thin, a requirement in order to observe strong quantum capacitance tuning, since no nucleation layers are needed, as would be the case for $HfO_2$ deposition on graphene [31].

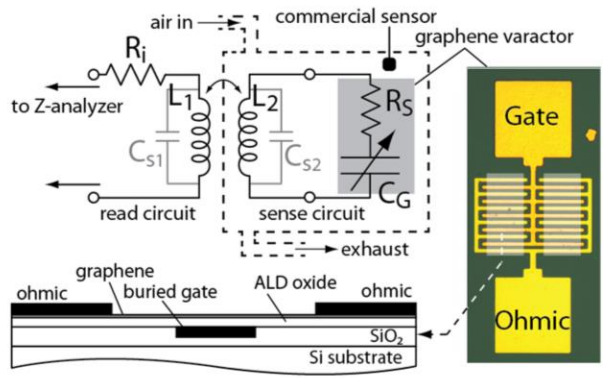

FIG. 2. Circuit diagram for the sensing system utilized in this work (top left) along with an optical micrograph of a typical varactor utilized for these experiments (right) and a cross-sectional schematic of the varactor structure (bottom left). Areas that include graphene have been highlighted with transparent white boxes in the micrograph. For the actual sensing experiments, five varactors similar to the one shown above were wire-bonded in parallel.

A diagram of the device design as well as an optical micrograph of a single graphene varactor are shown in Fig. 2.

### B. Humidity Sensing

Fig. 2 also shows a diagram of the flow cell geometry used for the humidity measurements as well as the circuit diagram for the coupled-inductor oscillator used for the wireless transduction. The relative humidity in the cell was controlled by mixing known flow rates of water-saturated and dry air (100% and ~0% relative humidity, respectively). Water-saturated air was produced by passing compressed air through a diffusing stone immersed in warm deionized water while dry air was produced by passing through a chamber packed with anhydrous calcium sulfate as a drying agent. To prevent condensed droplets of water from entering the sample chamber, a condensation trap was included in the water-saturated line immediately before mixing the wet and dry stream. Desired relative humidity values were achieved by carefully controlling the ratio of wet and dry air using valves and monitoring flow rate with rotameters inserted in each line. As an external calibrant, the relative humidity within the sample chamber was also monitored using an Electro-Tech Systems Model 514 humidity controller. In all of the measurements, at no time did condensation appear on the chip, which, on separate samples, was observed to abruptly change the resonant frequency.

After a 24 hour thermal bake at 380 K in vacuum to desorb water from the graphene surface, the sensor was immediately installed into the vapor chamber and the sensor inductor was aligned with a secondary read inductor (the read inductor did not include a Fe core) positioned on the exterior of the sample chamber. The read inductor was directly coupled to an Agilent 4291B impedance analyzer to measure the impedance and phase of the coupled inductor system.

To improve the accuracy of the quantitative fits, the inductances and self-capacitances of the sense and read inductors were independently determined using an Agilent 4291B impedance analyzer. The measured inductance values of the read and sense inductors were found to be $L_1$ = 1.16 μH and $L_2$ = 645 nH, respectively, with self-capacitances of $C_{S1}$ = 2.16 pF and $C_{S2}$ = 2.30 pF, respectively. These values were used when performing all quantitative fits for the wireless measurements.

### III. RESULTS AND DISCUSSION

#### A. Graphene Varactor Performance

Before testing, the chip was mounted on a printed circuit board and five varactors wire-bonded in parallel in order to increase the total capacitance. Prior to measurement, the mounted chip was baked at 380 K in vacuum to remove adsorbed water. Capacitance–Voltage ($C$–$V$) measurements were taken on the parallel wire-bonded varactors prior to removing from vacuum. The resulting 1 MHz $C$–$V$ curve is shown in Fig. 3. The characteristic quantum capacitance minimum is clearly observed just above the zero bias point. The capacitance tuning range ($C_{max}/C_{min}$) was found to be ~ 1.20. Fitting of the $C$–$V$ curve to a theoretical model [13] allowed for determination of the following device characteristics. The extracted equivalent oxide thickness ($EOT$) for the 8 nm-thick $HfO_2$ gate oxide was 2.52 nm (corresponding to a relative permittivity of 12.3). The fit also revealed a residual temperature, $T_0$, of 1500 K, where $T_0$ is related to the magnitude of the random potential disorder in the graphene. Furthermore, the area of the varactors was used as a fitting factor to account for tearing and delamination of the graphene in the active device area. The extracted value was $A$ = 7975 μm$^2$. Additional detail of the quantum capacitance fitting procedure is described in the appendix. It is important to note that the $C$–$V$ curve exhibits a steep slope near zero applied gate voltage. This condition is required to achieve high sensitivity during sensor operation.

The quality factor, $Q_{var}$, of the parallel varactors was also measured as a function of frequency, $f$, and these results are shown in the inset of Fig. 3. For the stand-alone varactors, $Q_{var}$ is defined as $1/2\pi f R_s C_G$, where $R_s$ is the series resistance and $C_G$ is the varactor capacitance. The relatively low frequency at which $Q_{var}$ rolls off indicates that $R_s$ is higher than would be expected given the graphene mobilities and contact resistances typically measured using our fabrication process. This excess series resistance is believed to be associated with graphene tearing at the edges of the gate electrode and is expected to be minimized using a more sophisticated planarization process, such as chemical-mechanical polishing. Nevertheless, the observed $Q_{var}$ value was sufficient to perform the wireless sensing measurements described in the next section.

#### B. Wireless Humidity Sensing

To make a basic demonstration of the quantum capacitance-based sensing, the graphene varactor was tested as a humidity sensor. While many more technologically interesting analytes exist, water vapor sensing represents the simplest method to demonstrate the quantum capacitance-based transduction mechanism, which is the focus of this paper. While pristine graphene has been shown to be intrinsically insensitive to changes in relative humidity, the presence of polymeric residues resulting from the transfer and subsequent lithography of graphene has been shown to impart sensitivity to the graphene [15]. Moreover, the presence of defect sites and crystalline boundaries in CVD-grown graphene lead to oxygen-containing moieties on the graphene [32]. Such functionalities have previously been suggested as active sites which lead to the sensitivity of CVD graphene-based devices [33].

In the intended mode of operation, adsorbed water on the graphene surface increases the hole concentration in the already slightly p-type graphene [34]. The increasing hole concentration shifts the Fermi-level further from the Dirac energy, increasing the capacitance and thus decreasing the resonant frequency of the LRC circuit.

As an initial test of the sensors, $\theta_z$ vs. $f$ for the external inductor was measured first in the dry condition, then in the humid condition and again in dry air. Here, the "dry" state corresponds to ~ 1% RH, with the "humid" state occurring at RH ~ 97%. In this initial test, the chamber RH was allowed to fully equilibrate under dry conditions before the measurements were taken and $\theta_z$ vs $f$ recorded at several time increments while changing RH. The $\theta_z$ vs $f$ curves taken at dry and humid conditions are shown in Fig. 4a. The minimum phase dip, which corresponds to the resonant frequency of the LRC sensor circuit, is clearly seen to shift to lower values under humid conditions and then returns to its original value in dry air. Fig. 4b also shows the measured impedance magnitude for the dry and humid conditions. To demonstrate the time response of the quantum capacitance sensor, a plot of the resonant frequency as a function of time is shown in Fig. 4c, while the RH vs. time plot measured using a commercial humidity sensor is shown in Fig. 4d. In Fig. 4c, two profiles

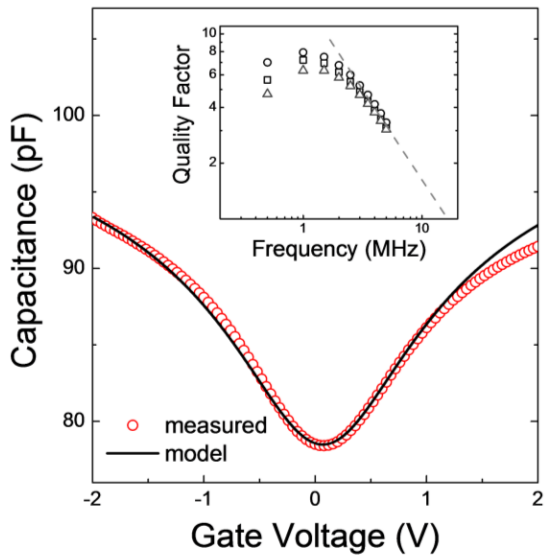

FIG. 3. Measured and modeled capacitance vs. voltage characteristics of graphene varactor utilized for sensing experiments. The device consisted of 5 multi-finger graphene varactors wire bonded in parallel, with aggregate area estimated to be 7975 μm$^2$. The measurement frequency is 1 MHz. Inset: log-log plot of quality factor vs. frequency for the graphene varactors.

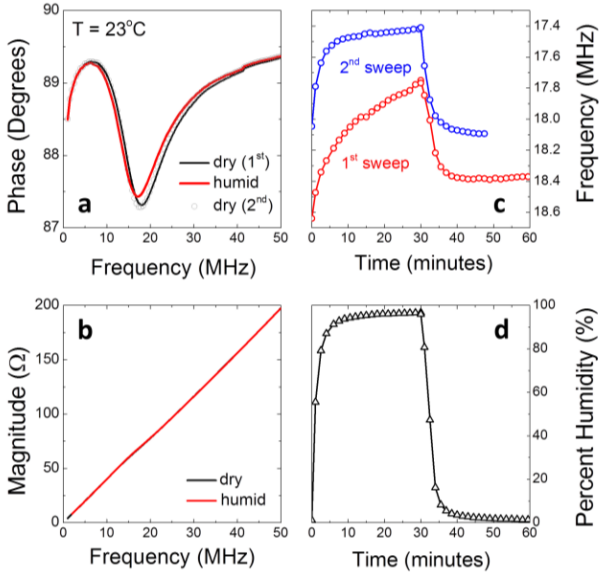

FIG. 4. (a) Plot of external inductor impedance phase versus frequency for successive measurements in dry (1% RH), humid (97% RH) and dry air. (b) Plot of external inductor impedance magnitude for the first two dry and humid conditions in (a). (c) Resonant frequency shift vs. time for two successive measurements where the RH was switched from the dry to humid states. The first profile was taken immediately after baking out in vacuum, while the second profile was performed after cycling the sensor between dry and humid conditions numerous times. (d) RH vs. time plot measured using a commercial humidity sensor.

## C. Effect of Concentration Cycling

To characterize both the concentration response and reproducibility of the sensor, three concentration-dependent resonant frequency profiles were measured, as summarized in Fig. 5a. The first profile followed a decreasing sequence from high to low concentration (Fig. 5b). Between each concentration, the humidity was brought to a minimum (~2% relative humidity) to track hysteretic behavior. The second profile followed an increasing sequence from low to high concentration (Fig. 5b). Finally, the third profile was taken such that the humidity concentration target was randomized, and the concentration sequence for this measurement is shown in Fig. 5c. It is notable that the resonant frequency shift as a function of concentration is roughly linear regardless of sweep direction, though a slight difference in the slopes are plotted which correspond to successive measurements of the graphene sensor on different days. The first profile was taken immediately after baking out in vacuum, while the second profile was performed after cycling the sensor between dry and humid conditions numerous times. In the first plot, it can be seen that the resonant frequency does not return to its original value after humidity cycling, but that the second curve does.

The second profile showed a net downward shift in resonant frequency of approximately 400 kHz with respect to the initial humidity ramp. The time response of the resonant frequency follows an approximate exponential curvature, and has a time response that is nearly equal to the commercial humidity sensor. It is speculated that the improved response observed in the second profile is a result of "seasoning" of the graphene in which the first profile contains some amount of transients related to the freshly dehydrated surface that are later equilibrated after exposure to a humid environment. Specifically, the surface of the hafnium oxide gate dielectric is expected to become dehydrated during a vacuum bake-out. Upon exposure to humid atmosphere, this surface is expected to again become hydrated [35]. Our results indicate that equilibration of the sensor is largely complete after 24 hours of exposure to atmosphere. Additionally, the sensor shows a steady response after 30 minutes in the second humidity cycle, indicating that the transients involved in the first cycle have been largely eliminated. It is also noted that no measurable difference in the response time of the graphene quantum capacitance sensor and the commercial humidity sensor was observed.

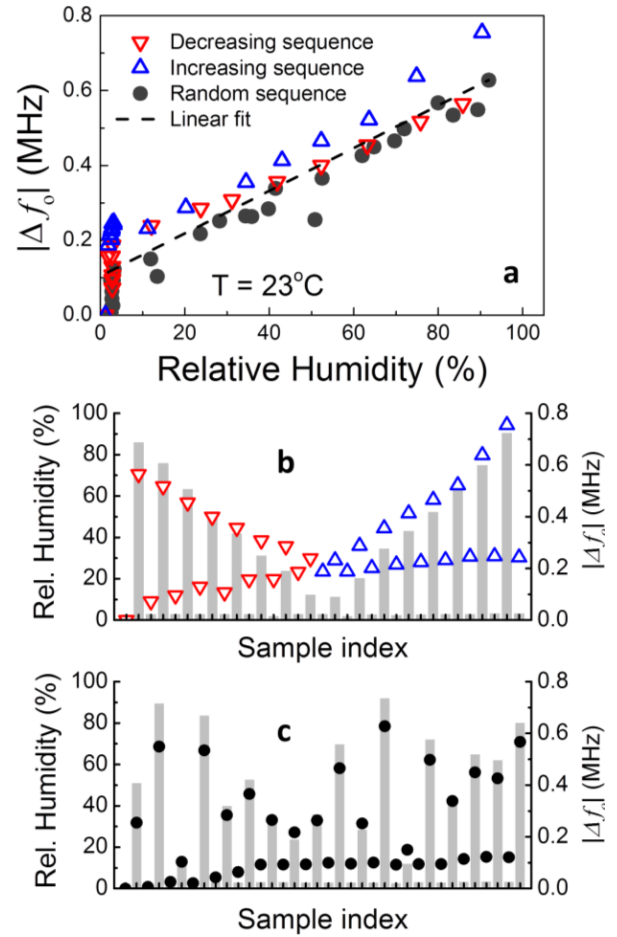

FIG. 5. (a) Dependence of resonant frequency shift vs. RH measured using three different concentration sequences: increasing, decreasing and random. The dashed line shows a linear fit including all three measurement sequences. (b) Measurement sequence for decreasing and increasing concentration-dependent measurements. (c) Measurement sequence for random concentration-dependent measurements. For all measurements in (b) and (c), the RH was cycled back to the dry condition between each concentration value. The measured RH values using a commercial humidity sensor are shown by the gray bars, while the resonant frequency shifts of the graphene sensor are depicted using the symbols.

corresponding to increasing and decreasing RH is apparent in Fig. 5a. Furthermore, we note that the slope of the frequency shift vs. concentration plot obtained from the randomized RH sequence is approximately the average of the slopes corresponding to the increasing and decreasing humidity sweeps. This indicates that a small but non-negligible hysteretic mechanism could still be at work that causes the frequency shift to be dependent on the direction of the concentration ramp.

It is interesting to note that although the results in Fig. 5a show a linear dependence of the frequency shift on humidity, such a functional dependence is not necessarily expected, as noted originally in reference [14]. Rather, the precise functional dependence is expected to depend upon numerous factors, including the interaction of the adsorbed molecules on the graphene surface, the precise shape of the $C$–$V$ profile and the initial "doping" in the graphene. In order to determine the precise operating conditions of our devices, we modeled the response of the sensors using the circuit impedance method described in reference [24] with a quantum capacitance model including random potential variations adapted from reference [13]. Our circuit model includes the effect of inductor self-resonance due to inter-winding capacitance.

### D. Equivalent Circuit Modeling

Fig. 6 shows the results of fitting the measured impedance phase data to the circuit model described in the appendix. Fig. 6a shows the measured phase dip under dry and humid conditions along with the modeled phase dip data. The only free fitting parameters were the sensor capacitance and resistance, the read inductor series resistance and the inductor coupling coefficients, while the coil inductance and self-capacitance values had been measured independently as described earlier. In total, eight relative humidity points were chosen for parameter extraction from the model which allowed the estimation of the change in resistance and capacitance of the sensor circuit as a function of RH. The extracted resistance and capacitance values vs. RH are shown in Fig. 6b and Fig. 6c, respectively. The capacitance is observed to decrease by roughly 10% over the range of vapor concentrations tested, while the resistance changed by < 1%. It is important to point out that if resistance changes were the primary transduction mechanism, these changes would mostly manifest as a change in the full-width half-maximum (FWHM) of phase dip signal, since the varactor resistance serves as the primary damping factor of the resonant circuit. Instead a frequency shift is observed, which is indicative of capacitance modulation. Therefore, these results provide firm evidence that the fundamental sensing mechanism involved in these sensors is in fact due to the quantum capacitance modulation of the graphene varactor. However, it should be noted that the resistance change extracted from the phase-dip measurements shows little change with increasing RH values, an unexpected result given previous studies on resistive graphene moisture sensors [15]. This discrepancy can be partially explained by the high series resistance in our devices, which would be expected to reduce the percent resistance change resulting from a shift in the carrier concentration relative to reference [15]. However, further study of the coincident resistance and capacitance changes in these sensors is still needed.

As a final demonstration of the quantum capacitance transduction mechanism, the known $C$–$V$ characteristics shown in Fig. 3 were used to extract the quantum capacitance vs. RH and these results are shown in Fig. 7. Using these values, it is observed that the humidity shifts the quantum capacitance between values of 3.5 $\mu$F/cm$^2$ and 4.9 $\mu$F/cm$^2$. This information could be extremely useful in understanding the fundamental properties of surface adsorption onto graphene since, unlike resistance-based sensors, the quantum capacitance sensor provides a method to directly link the adsorbed molecular concentration to carrier concentration changes.

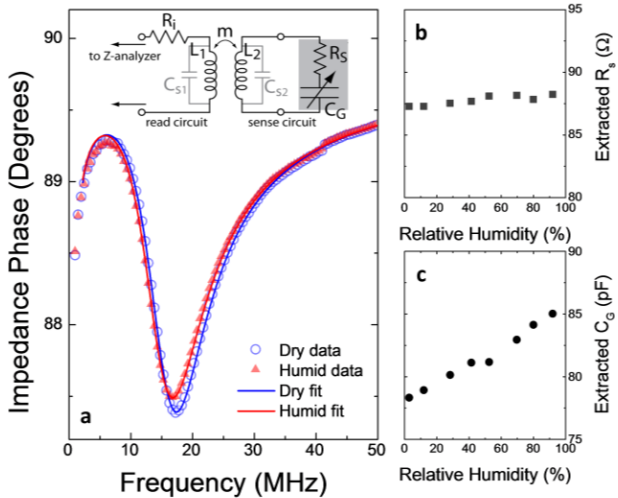

FIG. 6. (a) Measured phase dip under dry and humid condition along with the results of modeling using the equivalent circuit shown in the inset. The fitting parameters were the resistance and capacitance of the graphene varactor, the read inductor resistance and coupling coefficient between the two inductors. All other parameters were measured independently. Extracted (b) resistance and (c) capacitance of the graphene varactors vs. RH using the fitting procedure shown in (a).

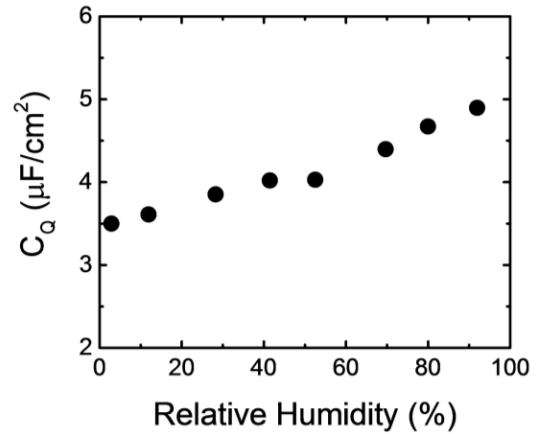

FIG. 7. Plot of quantum capacitance vs. RH extracted from the total capacitance vs. RH shown in Fig. 6 and the theoretical fit of the $C$–$V$ curve plotted in Fig. 3.

## IV. CONCLUSIONS

In conclusion, graphene vapor sensors that utilize the quantum capacitance effect as their principle of operation have been demonstrated. The sensors transduce a change in adsorbed water vapor concentration on the graphene surface, into a shift in the resonant frequency of a resonant oscillator circuit. The sensors show fast response to abrupt changes in the humidity and further show a monotonic frequency shift with relative humidity that is reversible and stable, particularly after conditioning using repetitive humidity cycling. Our results suggest that graphene quantum capacitance wireless sensors can be utilized to realize passive sensors for detection of a wide range of chemical and biological analytes, provided that appropriate surface functionalization approaches can be developed.

## V. APPENDIX

### A. Quantum Capacitance Model

The varactor $C$–$V$ characteristics in Fig. 3 were fit to a theoretical model assuming series connected oxide and quantum capacitances. For the quantum capacitance model, a fitting procedure has been established that takes into account the random potential fluctuations that can be particularly prominent in CVD-grown graphene. Using this model, the total varactor capacitance can be expressed as

$$C_G = A \cdot \left( \frac{1}{c_{ox}} + \frac{1}{c_Q} \right)^{-1} \tag{1}$$

where $A$ is the active area of the graphene, and $c_{ox}$ and $c_Q$ are the oxide capacitance and quantum capacitance per unit area, respectively. The oxide and quantum capacitance values can then be expressed as follows:

$$c_{ox} = \frac{3.9\varepsilon_0}{EOT} \tag{2}$$

and

$$c_Q = \frac{2}{\pi} \left( \frac{qkT_{eff}}{\hbar v_F} \right)^2 \ln\left( 2 + 2\cosh\left( \frac{E_F}{kT_{eff}} \right) \right) \tag{3}$$

Here, $\varepsilon_0$ is the permittivity of free space, $EOT$ is the equivalent oxide thickness of the dielectric between the metal gate electrode and the graphene, $q$ is the electronic charge, $k$ is Boltzmann's constant, $\hbar$ is the reduced Planck's constant, $v_F = 1.1 \times 10^6$ cm/s is the Fermi velocity, and $E_F$ is the Fermi energy relative to the Dirac point energy. $T_{eff}$ is the effective temperature, and is determined using:

$$T_{eff} = \sqrt{T_0^2 + T^2} \tag{4}$$

where $T$ is the sample temperature and $T_0$ is a fitting parameter intended to approximate the Dirac point "smearing" associated with random potential fluctuations [36].

### B. Impedance Model for Wireless Measurements

The basic principle of the phase-dip measurement is as follows. The frequency-dependent input impedance for the coupled readout and sensor circuit shown in Fig. 2, using the transformer equations for the inductively coupled circuit, is given as:

$$Z_{in} = Z_1 + \frac{\omega^2 m^2}{Z_2 + R_S + 1/j\omega C_G}, \tag{5}$$

where

$$Z_1 = R_i + \frac{j\omega L_1}{1 - \omega^2 L_1 C_{S1}}, \tag{6}$$

and

$$Z_2 = \frac{j\omega L_2}{1 - \omega^2 L_2 C_{S2}}. \tag{7}$$

In the above equations, $Z_1$ is the impedance of the read branch of the circuit, and $Z_2$ is the portion of the impedance of the sensor branch excluding the varactor elements. In addition, $\omega$ is the angular frequency, $L_x$ is the inductance of coil $x$, $C_x$ is the inter-winding capacitance of coil $x$, $R_i$ is the resistance of the read coil, $m = k(L_1 L_2)^{1/2}$ is the mutual inductance between the coils, and $k$ is the coupling coefficient. The varactor series resistance and capacitance are denoted by $R_S$ and $C_G$, respectively. When the sensor-side LRC circuit is at its resonant frequency, a plot of the phase of $Z_1$ vs. frequency has a minimum. Sensing occurs when the varactor capacitance varies in response to an external stimulus, which changes the resonant frequency, and therefore the value of the phase dip frequency. The fitting results are shown in Fig. 6, and for all fits, the values of $R_i$ and $k$ were used as free fitting parameters, where values of $R_i = 0.093$ Ω, $k = 0.16$ were determined in all cases.

**David A. Deen** (S'10–M'11) received the B.S. degree in Engineering Physics from the University of Oklahoma in 2005, and the M.S. and Ph.D. degrees in Electrical Engineering from the University of Notre Dame in 2011.

His early research interests involved III-Sb based spintronics. From 2008 to 2011 he was a research engineer for the Naval Research Laboratory, Washington D.C. In conjunction with his work at NRL, his graduate research focused on design, modeling, fabrication, and analysis of III-Nitride semiconductor devices for high frequency/power applications. From 2012 to 2013 he was a Post-Doctoral Research Associate with the Department of Electrical and Computer Engineering, University of Minnesota, Minneapolis, MN where he conducted research on graphene devices and their implementation in passive electronic circuits. He is currently a senior R&D Engineer with Seagate Technology, Bloomington, MN.

**Eric J. Olson** received the B.S. degree in chemistry from South Dakota School of Mines and Technology, Rapid City, in 2007 and the M.S. and Ph.D. degrees in chemistry from the University of Minnesota-Twin Cities in 2009 and 2012, respectively.

He is currently a Post-Doctoral Researcher Associate with the Department of Electrical and Computer Engineering, University of Minnesota, Minneapolis, MN. His research interests include electrochemistry and selective glucose sensing. Dr. Olson was a recipient of the Undergraduate Award in Analytical Chemistry from the American Chemical Society, the Krogh Fellowship from the University of Minnesota-Twin Cities, and the Award for Outstanding Achievement at the Chemistry Graduate Research Symposium in 2010.

**Mona A. Ebrish** (S'11) received the B.S. degree in electrical engineering from the University of Tripoli, Libya in 2007 and M.S. degree in electrical engineering from the University of Minnesota-Twin Cities in 2011. She is currently pursuing the Ph.D. degree in electrical engineering from the University of Minnesota-Twin Cities.

Her research interests include studying the physics of the interface between 2-D materials and their 3-D surroundings and exploring new applications for graphene devices, including remote sensing and in vivo applications. Ms. Ebrish received the Fulbright Fellowship in 2009 and is also a member of the Materials Research Society.

**Steven J. Koester** (M'96–SM'02) received the B.S.E.E and M.S.E.E. degrees from the University of Notre Dame, Notre Dame, IN, in 1989 and 1991, respectively, and the Ph.D. degree, in 1995, from the University of California, Santa Barbara, where his research involved the study of quantum transport in InAs quasi-1-D structures.

He has been a Professor of Electrical and Computer Engineering in the College of Science and Engineering at the University of Minnesota, in Minneapolis, MN since 2010. Prior to joining the University of Minnesota, he was a Research Staff Member with the T. J. Watson Research Center, IBM Research Division, Yorktown Heights, NY where his work involved Si/SiGe devices and materials, high-speed Ge photodetectors, and III-V MOSFETs. His most recent position at IBM was Manager of Exploratory Technology where his team investigated novel device and integration solutions for post-22-nm node CMOS technology.

Dr. Koester's current research involves investigations into the device applications of graphene, including novel sensors, spintronics, and optoelectronic devices. He has authored or coauthored more than 160 technical publications and conference presentations, and is the holder of 46 U.S. patents. He was the general chair of the 2009 Device Research Conference and is currently an associate editor of *IEEE Electron Device Letters*.